# Effect of higher borides and inhomogeneity of oxygen distribution on critical current density of undoped and doped magnesium diboride

**T A Prikhna**[1], **W Gawalek**[2], **V M Tkach**[1], **N I Danilenko**[3], **Ya M Savchuk**[1],
**S N Dub**[1], **V E Moshchil**[1], **A V Kozyrev**[1], **N V Sergienko**[1], **M Wendt**[2],
**V S Melnikov**[1], **J Dellith**[2], **H Weber**[4], **M Eisterer**[4], **Ch Schmidt**[2],
**T Habisreuther**[2], **D Litzkendorf**[2], **J Vajda**[5], **A P Shapovalov**,[1] **V Sokolovsky**[6],
**P A Nagorny**[1], **V B Sverdun**[1], **J Kosa**[5], **F Karau**[7], **A V Starostina**[1]

[1] Institute for Superhard Materials of the National Academy of Sciences of Ukraine, 2, Avtozavodskaya Str., Kiev 04074, Ukraine

[2] Institut für Photonische Technologien, Albert-Einstein-Strasse 9, Jena, D-07745, Germany

[3] Institute for Problems of Materialscience of the National Academy of Sciences of Ukraine,3, Krzhyzhanivsky Str., Kiev 03142, Ukraine

[4] Atomic Institute of the Austrian Universities, 1020 Vienna, Austria

[5] Budapest University of Technology and Economics, Budapest, Hungary 1111 Budapest, Egry Jozsef u. 18. Hungary

[6] Ben-Gurion University of the Negev, P.O.B. 653, Beer-Sheva 84105 Israel

[7] H.C. Starck GmbH, Goslar 38642, Germany

E-mail: prikhna@iptelecom.net.ua, prikhna@mail.ru

**Abstract**. The effect of doping with Ti, Ta, SiC in complex with synthesis temperature on the amount and distribution of structural inhomogeneities in $MgB_2$ matrix of high-pressure-synthesized-materials (2 GPa) which can influence pining: higher borides ($MgB_{12}$) and oxygen-enriched Mg-B-O inclusions, was established and a mechanism of doping effect on $j_c$ increase different from the generally accepted was proposed. Near theoretically dense SiC-doped material exhibited $j_c = 10^6$ A/cm$^2$ in 1T field and $H_{irr}$ =8.5 T at 20 K. The highest $j_c$ in fields above 9, 6, and 4 T at 10, 20, and 25 K, respectively, was demonstrated by materials synthesized at 2 GPa, 600 °C from Mg and B without additions (at 20 K $j_c$= 10$^2$ A/cm$^2$ in 10 T field). Materials synthesized from Mg and B taken up to 1:20 ratio were superconductive. The highest $jc$ (6×10$^4$ A/cm$^2$ at 20 K in zero field, $H_{irr}$= 5 T) and the amount of SC phase (95.3% of shielding fraction), $T_c$ being 37 K were demonstrated by materials having near $MgB_{12}$ composition of the matrix. The materials with $MgB_{12}$ matrix had a doubled microhardness of that with $MgB_2$ matrix (25±1.1 GPa and 13.08±1.07 GPa, at a load of 4.9 N, respectively).

## 1. Introduction

Synthesis and sintering of $MgB_2$ under high pressure conditions allow one to produce near theoretically dense nanostructural material with a good connectivity between grains and a high critical current density, $j_c$, for a short time. High pressure (1-2 GPa) suppresses the magnesium from being evaporated and prevents grain growth (the average sizes of crystallites estimated by X-ray did not surpass 15-37 nm) [1]. The progress in engineering of high-pressure apparatuses made it possible to synthesize large blocks (in our case above 60 mm in diameter), which makes this method interesting not only from the theoretical point of view.

---

[1] Institute for Superhard Materials of the NASU, 2, Avtozavodskaya Str., Kiev 04074, Ukraine.

In 2005 the first SC motor of zebra-type rotor was constructed using high pressure-synthesized (HPS) $MgB_2$. The tests showed its competitiveness at liquid hydrogen temperatures (15-20 K) with motor based on MT-YBCO, besides the high pressure-synthesized $MgB_2$ is two-three times lower in cost and easier in production than MT-YBCO [2]. The first investigations of high pressure-synthesized $MgB_2$ rings for inductive fault-current limiters in the Ben-Gurion University of the Negev (by V. Meerovich and V. Sokolovsky) showed encouraging results: the transition is very fast, cylinders can provide higher impedance change at the transition from nominal to limiting regime.

Due to the comparatively large coherent length (1.6-12 nm) pining centres in $MgB_2$ can be grain boundaries and nanosized inclusions. Because of this, high $j_c$ can be attained in nanocrystalline material and superconducting (SC) properties can be improved by alloying with nanosized properly distributed grains of additives. The atomic resolution study of oxygen incorporation into the bulk $MgB_2$ [3] shows that 20-100 nm sized precipitates of Mg(B,O) are formed by ordered substitution of oxygen atoms to boron lattice sites, while the basic bulk $MgB_2$ crystal structure and orientation are preserved. The periodicity of the oxygen ordering is dictated by the oxygen concentration in the precipitates and primarily occurs in the (010) plane [3]. The presence of these precipitates correlates well with an improved critical current density and superconducting transition behavior, implying that they act as pinning centres [3].

Our previous study has established that superconductive properties of high pressure manufactured $MgB_2$ depend on the amount, size and distribution of higher borides inclusions with near $MgB_{12}$ stoichiometry (the finer the $MgB_{12}$ grains and the larger amount of them, the higher $j_c$) [4]. Besides, it has been shown that additions of Ti, Ta, and Zr can improve critical current density $j_c$, upper critical field, $B_{c2}$, and irreversibility line, IL, of high-pressure high-temperature synthesized $MgB_2$ (Fig. 1), but we did not observe the same mechanism as it was described for magnesium diboride obtained under normal pressure conditions [6]. In case of adding Ti or Zr the improvement in critical current density in materials synthesized at ambient pressure was usually explained by the formation of $TiB_2$ or $ZrB_2$ thin layers or inclusions at grain boundaries of $MgB_2$ that increase the number of pinning centres, which is ascribed to a $j_c$ improvement caused by doping with these elements [6]. In high-pressure synthesized $MgB_2$, the Ti-, Zr- or Ta-containing inclusions are rather coarse (Fig. 2) and randomly distributed in the matrix material to be pinning centres by themselves or to refine the $MgB_2$ structure. Under high pressure Zr, Ti or Ta absorb impurity hydrogen (the source of which can be materials of high-pressure cell surrounded the sample during synthesis) to form $TiH_{1.94}$, $ZrH_2$, or $Ta_2H$ (Fig. 3a, b, c), thus preventing the harmful (for $j_c$) $MgH_2$ (Fig. 3d) impurity phase from appearing and hydrogen from being introduced into the material structure. Besides, it has been observed that the presence, for example, of Ti or Ta promotes the formation of $MgB_{12}$ inclusions, which positively affect pinning in $MgB_2$-based materials (Table. 1), while the appearance of $ZrB_2$ in the structure does not affect the $j_c$ of HPS $MgB_2$-based ceramics (Fig. 4). Ti- and Zr-doped HPS-synthesized materials demonstrated extremely high upper critical fields (Figs. 1c, d) and rather high $j_c$, which practically were not increased by irradiation [5]. The study of high pressure-synthesized material from Mg and B with 10% of Ti addition (the $j_c$ dependences, structure and X-ray pattern of which shown in Figs. 1, 2, 3b) by SIMS chemical analysis using a Cameca NanoSIMS 50 with a Cs+ primary ion beam (with possibility to map the hydrogen distribution in the specimens) and by Electron Microprobe analysis (EPMA) using a JEOL JXA 8800 Superprobe as well as by transmission electron microscopy on powdered samples dispersed on a lacy carbon film using a JEOL 3000F HR-TEM fitted with an energy dispersive X-ray (EDX) detector has shown that the only Ti-containing phase in the material was $TiH_{1.924}$ [7] despite the fact that the enthalpy of titanium oxides formation is lower than that of $TiH_2$.

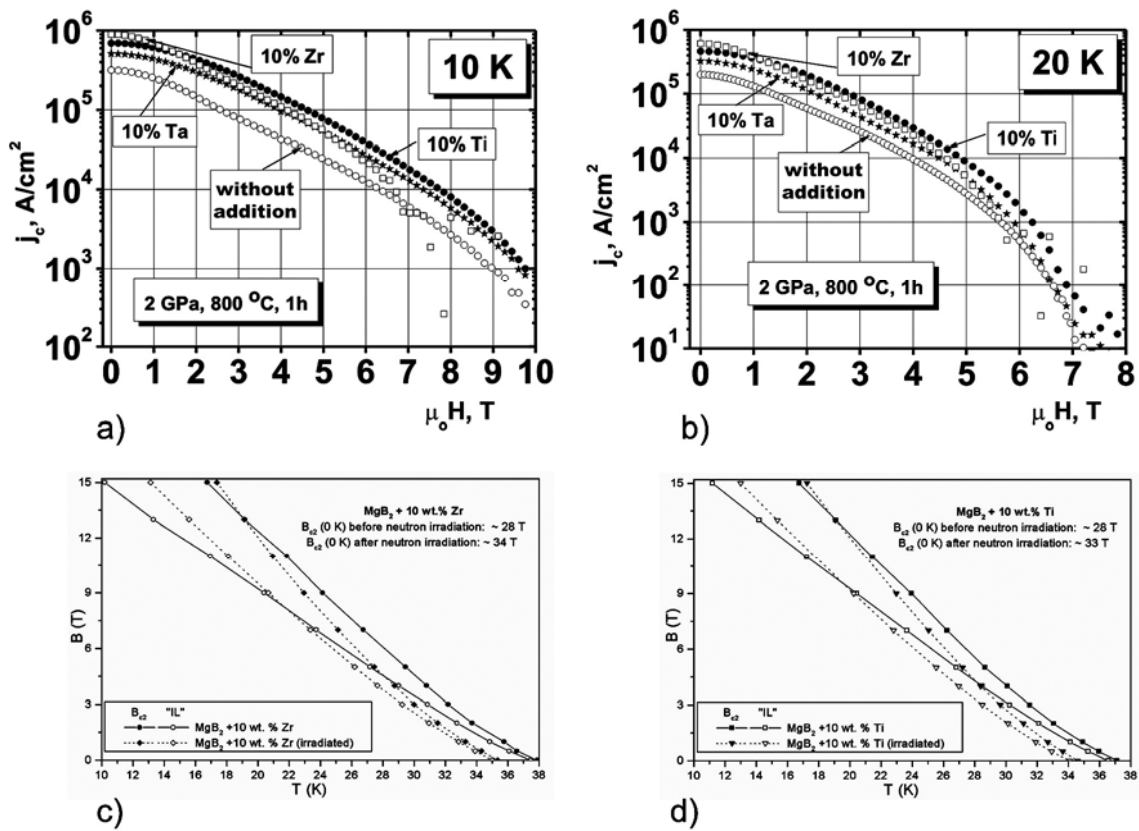

**Figure 1.** (a, b) Dependences of critical current density, $j_c$, on magnetic fields, $\mu_o H$ for HPS-materials at 2 GPa, 800 °C for 1 h prepared from Mg:B(IV)=1:2 without additions and with additions of 10 wt% of Ti, Ta and Zr at 10 K (a) and 20 K (b); (c, d) dependences of upper critical field, Bc$_2$, and irreversible line, IL, from temperature for HPS-materials at 2 GPa, 800 °C for 1 h prepared from Mg:B(IV)=1:2 with 10 wt% of Zr (c) and 10 wt% of Ti ( before and after irradiation [5]).

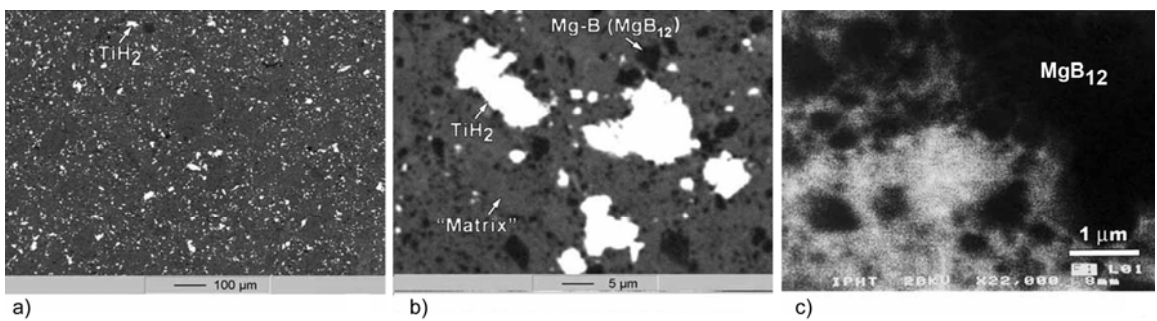

**Figure 2.** Structures (a, b, c) of high-pressure synthesized material from Mg:B(IV)=1:2 with 10 wt% of Ti at 2 GPa, 800 °C for 1 h at different magnifications (obtained using SEM, "composition" or backscattering electron images, in which the heavier element, the brighter it looks).

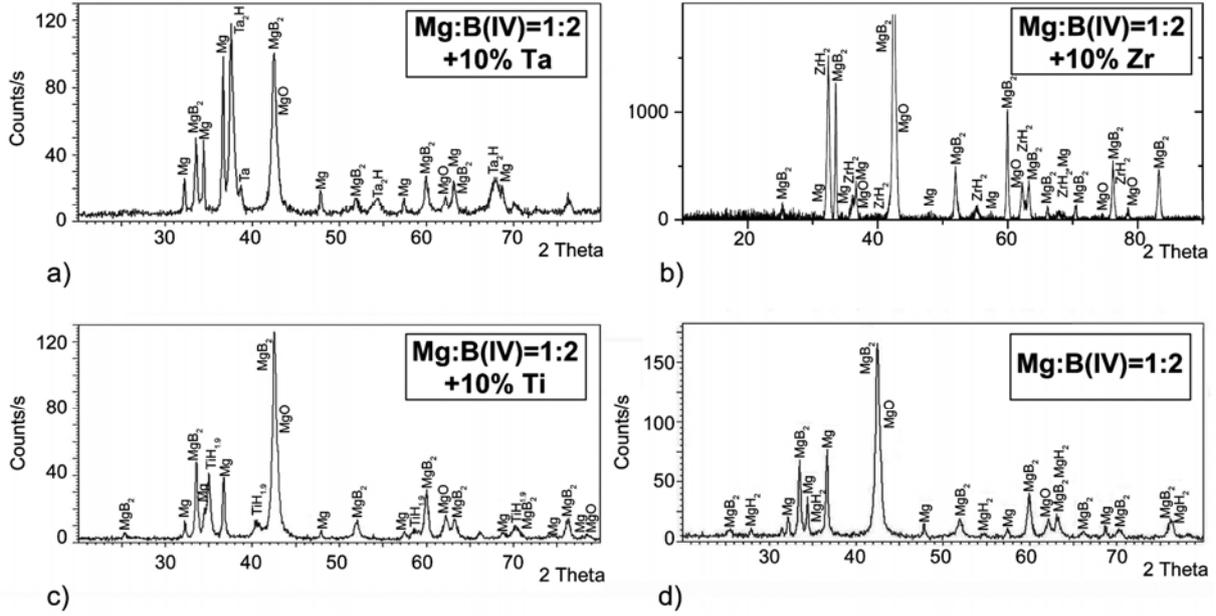

**Figure 3.** X-ray patterns of HPS-materials from Mg:B(IV)=1:2 at 2 GPa, 800 °C for 1h with 10 wt% of Ta (a), Ti (b), Zr (c) and without additives (d).

**Table 1.** Critical current density, $j_c$, vs. relative amount, N, of "black inclusions" with near $MgB_{12}$ stoichiometry in high-pressure synthesized samples from Mg and B(IV) taken in $MgB_2$ stoichiometry (or in 1:2 ratio).

| Manufacturing parameters: pressure, P, temperature, T, holding time, τ, | Name of addition, and its amount, wt% | $j_c$ in 1T, at 20 K kA/cm² | N, % [a] |
|---|---|---|---|
| P=2 GPa, T=800 °C, τ=1h | Ta, 10% | 240 | 12.5 |
| | Ti, 10% | 360 | 14 |
| | without | 131 | 10.8 |

[a] The amount of the "black" inclusions, $N$, was calculated as a ratio of the area occupied by the "black" inclusions at the COMPO image obtained at 1600x magnification to the total area of this image.

Here we discuss the appearance of structural inhomogeneities such as Mg-B-O inclusions and grains of higher borides in HPS $MgB_2$-based materials in connection with synthesis temperature and different additions (Ti, Ta or SiC) and their effect on SC properties of the materials. The SC behavior of HPS materials prepared from mixtures with enriched amount of boron (up to Mg:B=1:20) as compare to $MgB_2$ stoichiometry and the possibility to get a high critical current density in the materials with near $MgB_{12}$ composition of the matrix are under consideration.

## 2. Experimental

Samples were high-pressure (2 GPa) synthesized at 600 – 1050 °C from Mg and B in recessed-anvil high-pressure apparatuses [4] in contact with hexagonal BN. As initial materials we used powder of $MgB_2$ (H.C. Starck) with an average grain size of 10 μm and 0.8 % of O; several types of amorphous boron (H.C. Starck): B(I) 1.4 μm, 1.9% O, B(II) <5 μm 0.66 % O, B(III) 4 μm, 1.5 % O; B(IV) MaTecK 95-97% purity, 0.8 μm and 1.7 % of O, B(V) HyperTech(USA) produced; metal magnesium chips (Technical Specifications of Ukraine 48-10-93-88), powder of Mg(A) HyperTech (USA) produced, Ti (of size 1-3 μm, MaTecK, 99% purity), Ta (technical specifications 95-318-75, 1-3 μm) or SiC (200-800 nm, H.C. Starck). To produce MgB$_2$-based materials, metal magnesium chips and amorphous boron were taken in the stoichiometric ratio of MgB$_2$. To study the influence of Ti or SiC, powders were added to the stoichiometric MgB$_2$ mixture in amount of 10 wt%. The components were mixed and milled in a high-speed activator with steel balls for 1-3 min and then tabletized. To study the processes of higher borides formation, Mg and B (III) were taken in 1:4, 1:6, 1:7, 1:8, 1:10, 1:12, and 1:20 ratios and HPS at 800 and 1200 $^o$C at 2 and 4 GPa for 1h.

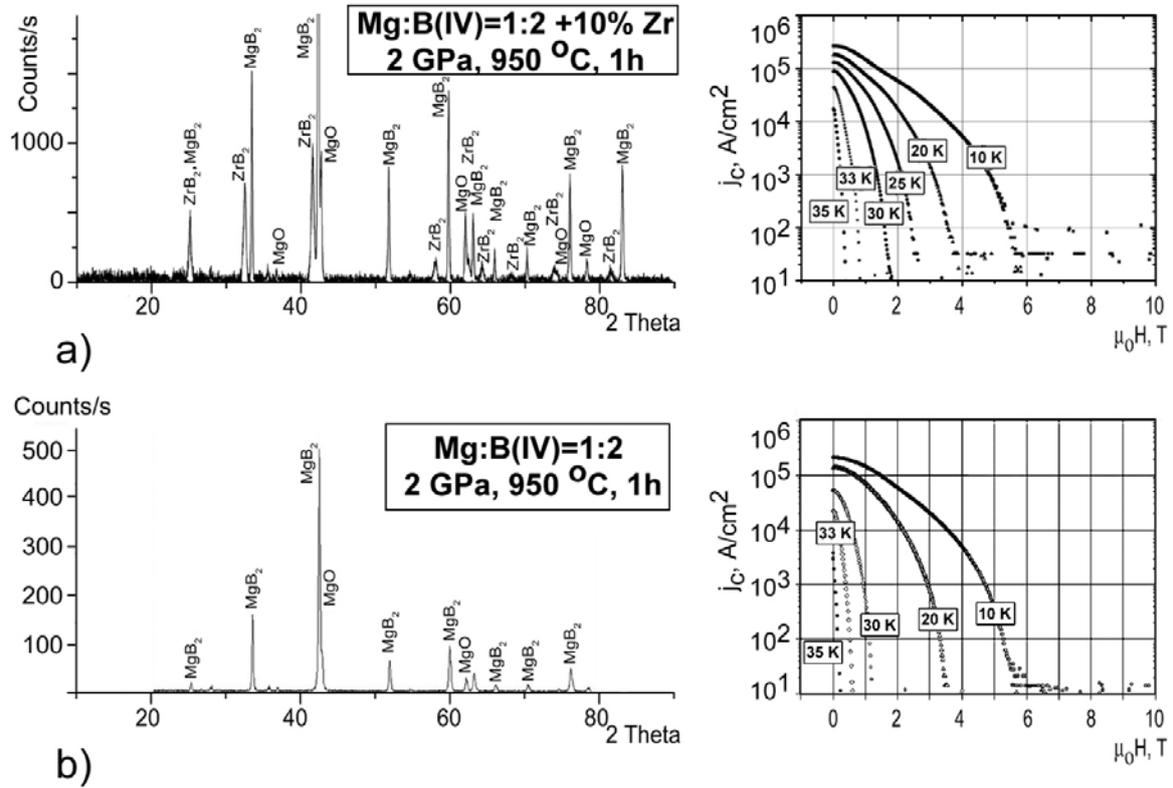

**Figure 4.** X-ray patterns and dependences of critical current density, $j_c$, on magnetic fields, $\mu_oH$, for HPS-materials at 2 GPa, 950 $^o$C for 1 h from Mg:B(IV) with 10 wt% Zr (a) and without additives(b).

The structure of the materials was analyzed using TEM, SEM, and X-ray diffraction. A scanning electron microscope ZEISS EVO 50XVP (resolution of 2 nm at 30 kV), equipped with: (1) an INCA 450 energy-dispersion analyzer of X-ray spectrums (OXFORD, England), using which the quantitative analysis from boron to uranium with a sensitivity of 0.1 wt % can be performed; probe 2 nm in diameter; (2) a HKL Canell 5 detector of backscattering electrons (OXFORD, England), which allows us to get (using the Kikuchi method) the diffraction reflections of electrons from 10-1000 nm areas and layers was employed. For SEM study JXA 88002 was used as well. The microstructure analysis on the nanometer scale was carried out using JEM-2100F TEM equipped with an Oxford INCA energy detector. Quantitative TEM-EDX analysis of boron was performed using the Oxford INCA energy program, 0.7 nm probe diameter.

The average crystallite sizes from line broadening in X-ray diffraction pattern were calculated by the standard program in accordance with the following:

$$\text{Crystallite size} = \frac{K \cdot \lambda}{W_{\text{size}} \cdot \cos\theta} \quad \text{with} \quad W_{\text{size}} = W_b - W_s \qquad (1)$$

where $W_{size}$ is the broadening caused by small crystallites; $W_b$ is the broadened profile width; $W_s$ is the standard profile width (0.08 °); $K$ is the form factor; $\lambda$ is the wavelength.

The values of $j_c$ were estimated by an Oxford Instruments 3001 vibrating sample magnetometer (VSM) using Bean's model. Hardness was measured using a Matsuzawa Mod. MXT-70 microhardness tester, $H_V$ (using a Vickers indenter) and Nano-Indenter II, $H_B$ (using a Berkovich indenter).

## 3. Results and discussions

There are many structural factors which can influence pinning and $j_c$ of $MgB_2$-based materials and it is very difficult to separate one factor from the others and to study the effect of the only definite one. Despite the comparatively simple $MgB_2$ crystallographic structure (hexagonal lattice) the structures of $MgB_2$-based materials are very complicated (see, for example Fig. 5 b). It is difficult to study correlations between the structure and $j_c$ first of all because of the nanosized grains of the main (matrix) phase and due to the fact that oxygen and hydrogen can easily incorporate into the material structure (the amount and distribution of them can essentially affect the SC characteristics). It is difficult to avoid the oxygen and hydrogen presence because $MgB_2$-based materials as well as initial B and Mg possess a high reactivity toward oxygen and hydrogen. Besides, the boron has a low X-ray atomic scattering factor [8] and even dealing with near theoretically dense materials it is not easy to estimate the amount of boron by energy-dispersive analysis even using high resolution TEM and SEM. We cannot detect the presence of $MgB_{12}$ in $MgB_2$ structure by X-ray due to weak diffraction signals (Figs. 5c, d) likely because $MgB_{12}$ phase is dispersed in the matrix material. Besides, the etalon X-ray pattern of $MgB_{12}$ is absent in the database and the literature data are contradictive: some researches considered that $MgB_{12}$ has hexagonal (or rhombohedral) lattice [9, 10] and the other ones that it is orthorhombic [11]. There still exists vagueness concerning appearance, composition and structure of higher boride phases ($MgB_4$, $MgB_6$ or $MgB_7$, $MgB_{12}$, $MgB_{16}$ or $MgB_{17}$ and $MgB_{20}$). Figures 5a, b, e can be the illustration of $MgB_{12}$ effect on $j_c$. The material sintered from $MgB_2$ contained much less grains with near $MgB_{12}$ stoichiometry (than that synthesized from Mg:B(I)=1:2 (Fig. 5b), which may affect $j_c$ (Fig. 5e), but the presence of this phase is not reflected in the X-ray pattern (Figs. 5c, d).

Our attempts to find correlations between the average grain size, oxygen content of the initial materials (B or $MgB_2$ powders) and the amount of oxygen in high-pressure-synthesized (so called *in-situ*) or sintered (so called *ex-situ*) magnesium diboride as well as with $j_c$ of these materials failed (Figs. 6, 7, Table 2). As is seen from Fig. 6 there is no correspondence between the oxygen content in starting boron or $MgB_2$, its amount in the material obtained and $j_c$, for example, at 20 K in 1T field. The average crystallite sizes as were shown by the calculations from the line broadening in the X-ray diffraction pattern increased with temperature (Table 2) especially for *in-situ* produced $MgB_2$, but the variation of $j_c$ does not correspond to variation in grain sizes. In low magnetic fields the critical current density of the *in-situ* prepared $MgB_2$ increases with increasing synthesis temperature from 800 °C to 1000 °C or increasing crystallite sizes from 15 to 37 nm, but in high or higher magnetic fields the opposite tendency is observed (Fig. 7). Besides, the values of $j_c$ of *in-situ* synthesized at 900 °C material with intermediate sizes of crystallites of 21 nm (Table 2) demonstrated the values $j_c$ lover than those of synthesized at 800 °C in investigated area (at temperatures from 10 to 35 K and magnetic fields 0 - 10 T).

Up to now the highest $j_c$ for high pressure-synthesized $MgB_2$-based materials in magnetic fields up to 9, 6, and 4.2 T at 10, 20, and 25 K, respectively were exhibited materials prepared at 2 GPa, 1050 °C for 1 h (Fig. 8c, solid symbols) from Mg and B(II) with 10 % of SiC (200-800 nm). But at higher temperatures, 30-35 K, up to 2-1 T fields the highest $j_c$ were for material synthesized from the same boron B(II) under the same conditions but without additives (Fig. 8d, solid symbols). The highest $j_c$ in magnetic fields higher than 9 T, 6 T, and 4.2 T at 10 K, 20 K, and 25 K were exhibited by materials HP-synthesized only at 600 °C from HyperTech produced Mg(A) and B(V) also without additives (Fig. 8d, open symbols). It should be mentioned that the last-mentioned material at 20 K showed $j_c = 10^2$ A/cm$^2$ in 10 T field, which is 100 times higher than that reported for material with SiC additive in [12]: "The irreversibility field ($H_{irr}$) for the SiC doped sample reached the benchmarking value of 10 T at 20 K, exceeding that of NbTi at 4.2 K".

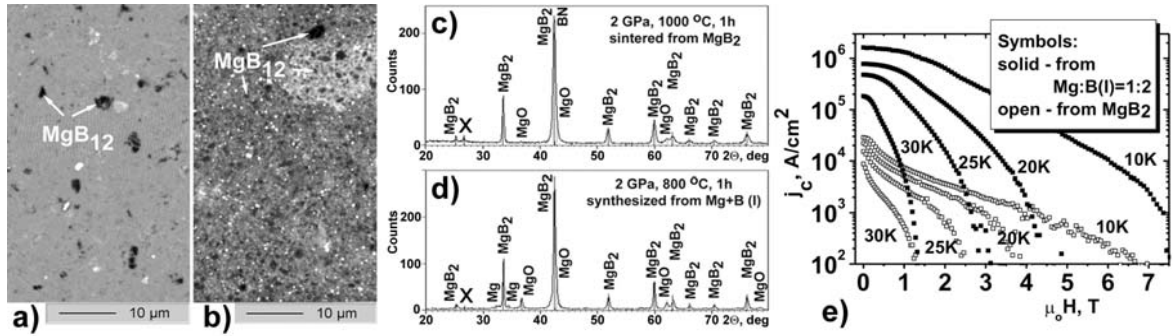

**Figure 5.** (a), (b) structures of the samples obtained by SEM in COMPOsitional contrast: (a) sintered from $MgB_2$ at 2 GPa, 1000 °C, 1 h; bright small inclusions in Fig. 5b seem to be inclusions because of milling (containing O, Zr and Nb, possibly $ZrO_2$), which were in the initial boron; bright bigger inclusions are unreacted Mg; (b) synthesized from Mg and B taken into 1:2 ratio at 2 GPa, 800 °C; (c), (d) – X-ray patterns of the samples shown in Figs. 5a, b; (e) dependences of critical current density, $j_c$, on magnetic fields, $\mu_o H$, at different temperatures of the samples shown in Figs. 1a, b: open symbols – sintered from $MgB_2$ material and solid symbols - synthesized from Mg and B (I) taken into 1:2 ratio.

As it is pointed out in [12] "a comparative study of pure, SiC, and C doped $MgB_2$ wires has revealed that the SiC doping allowed C substitution and $MgB_2$ formation to take place simultaneously at low temperatures. C substitution enhances $H_{c2}$, while the defects, small grain size, and nanoinclusions induced by C incorporation and low-temperature processing are responsible for the improvement in $J_c$". In the case of HPS materials we failed to improve $j_c$ using nano SiC additive with particle sizes of 20-30 nm as well as by adding SiC with particle sizes of 200-400 nm (Fig. 9). The only some improvement of $j_c$ was observed at 10 and 20 K in high magnetic fields (above 7 and 4 T, respectively) with nano SiC (20-30 nm) doping (Fig. 9a). The comparison of X-ray patterns (Figs. 8a, 9a-c) as well as SEM study witnessed that the improvement in $j_c$ in the case of HPS materials was obtained in the case when coarse-grained SiC of 200-800 nm was used and it was no notable interaction between SiC and $MgB_2$ (Fig. 8a). But in the case when this interaction took place (Figs. 9a, b) there were usually no improvements or a dramatic reduction of $j_c$ at 30-35 K even occurred. According to our results the pinning centers in SiC doped $MgB_2$ which influenced $j_c$ increase (compare Figs. 8b, d) may be SiC grains by themselves, inclusions of higher borides ($MgB_{12}$) and Mg-B-O inclusions enriched with oxygen as compare to oxygen content in the matrix phase (Figs. 10 g, h). The better SC properties can be reached when there is no notable reaction between SiC and $MgB_2$. So, it is most likely that the improvement in $j_c$ in HPS SiC-doped material is due to the formation of nanostructural inhomogeneties in $MgB_2$ matrix and presence of small SiC grains than due to carbon incorporation in $MgB_2$ crystal structure.

Figures 10 a-i show the structures of the materials high pressure synthesized from magnesium and different boron types taken in $MgB_2$ stoichiometry at 600, 800, and 1050 °C without and with Ti, Ta, and SiC additives. The matrix of all samples contained Mg and B in near-$MgB_2$ stoichiometry and some oxygen (5-10 wt%), because of this it was marked as Mg-B-O. Oxygen in $MgB_{12}$ inclusions was practically absent, 0.2-1.2 wt%. Structural observations showed that at a synthesis temperature of about 800 °C oxygen is comparatively homogeneously distributed in the matrix material (Figs. 10a, c), but as the synthesis temperature increases, the distribution of oxygen in the matrix is less homogeneous: the oxygen-enriched (as compared to the amount of oxygen in the matrix) areas form (Figs. 2b, 10b) and with addition of Ta or Ti such areas are transformed into separate Mg-B-O inclusions (light or white inclusions in Figs. 10 d-h). All materials also contain black inclusions with

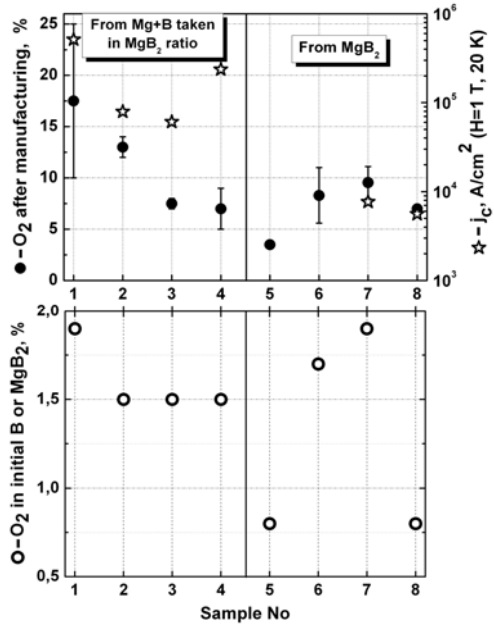

**Figure 6.** The amount of oxygen in the initial boron or MgB$_2$ powder (o) and in the prepared (•) materials (synthesized from Mg and B taken in 1:2 ratio or sintered from MgB$_2$) and critical current density (★) in 1 T field at 20 K estimated using VSM in different samples manufactured at 2 GPa for 1h at 800 °C (Nos 1, 2, 5), 900 °C (Nos 3, 6, 7) and 1000 °C (Nos 4, 8). The shape of hysteresis loops of samples 5 and 8 witnessed the absence of connectivity between the superconductive grains in the materials, thus as a whole the materials were not superconductive. The sample 1 prepared from boron with average grain size of 1.4 μm, samples 2, 3, 4 from 4 μm boron and samples 5 and 8 from 10 μm, sample 6 from 9 μm, sample 7 from 1.4 μm MgB$_2$.

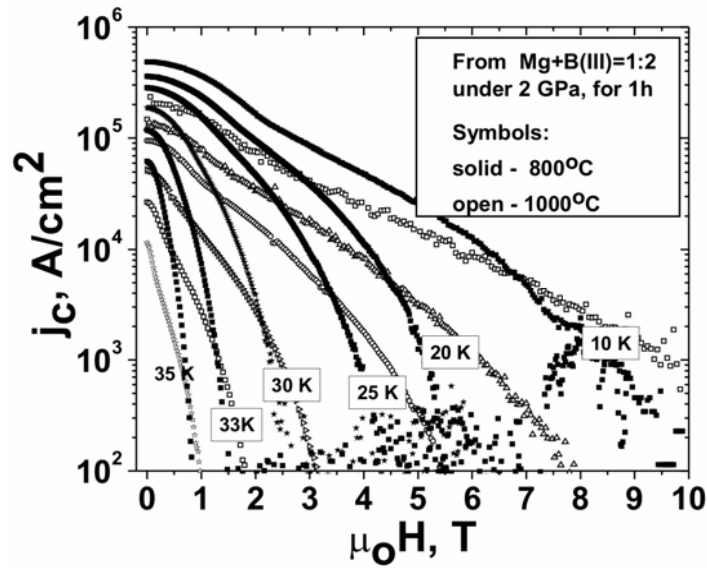

**Figure 7.** Dependences of critical current density, $j_c$, at different temperatures on magnetic fields, $\mu_o H$, for HPS-materials at 2 GPa for 1h from Mg:B(III)=1:2 at 800 °C (open symbols) and 1000 °C (solid symbols).

**Table 2.** Critical current density, $j_c$, vs. relative average grain size of crystallites of high-pressure sintered from $MgB_2$ and synthesized from Mg and B(III) taken in 1:2 ratio materials.

| HPS under 2 GPa for 1 h at $T_s$, °C | average crystal size | lattice parameters a (nm) | c (nm) | $j_c$, kA/cm² at 10 K 0 T | 1 T | $j_c$, kA/cm² at 20 K 0 T | 1 T |
|---|---|---|---|---|---|---|---|
| From $MgB_2$ | | | | | | | |
| 700 | 19.7 nm | 0.30805 | 0.35188 | - | - | - | - |
| 800 | 18.8 nm | 0.30822 | 0.35212 | - | - | - | - |
| 900 | 18.5 nm | 0.30820 | 0.35208 | 56 | 14 | 36 | 8 |
| 1000 | 25.0 nm | 0.30797 | 0.35200 | 28 | 8 | 19 | 5 |
| From Mg and B (III) in 1:2 ratio | | | | | | | |
| 800 | 15.0 nm | 0.30747 | 0.35188 | 245 | 142 | 138 | 79 |
| 900 | 21.0 nm | 0.30819 | 0.35174 | 205 | 136 | 128 | 61 |
| 1000 | 37.0 nm | 0.30808 | 0.35192 | 485 | 364 | 360 | 237 |

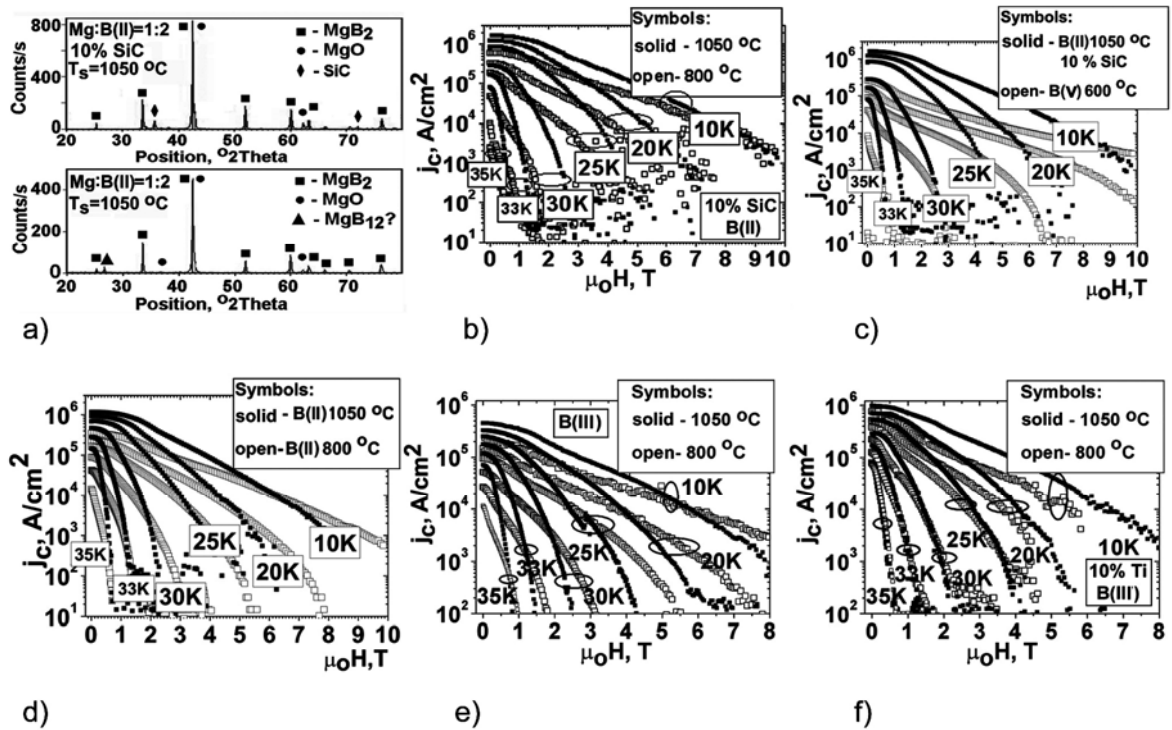

**Figure 8.** (a) X-ray patterns of materials synthesized from Mg and B(II) taken into 1:2 ratio at 2 GPa, 1050 °C for 1h with SiC additive (upper picture) and without additive (lower picture); (b-f) dependences of critical current density, $j_c$, at different temperatures on magnetic field, $\mu_o H$, for HPS-materials at 2 GPa for 1h: (b) from Mg:B(II)=1:2 with 10% SiC (200-800 nm) at 800 °C (open symbols) and 1050 °C (solid symbols), (c) from Mg:B(II)=1:2 with 10% SiC (200-800 nm) at 1050 °C (solid symbols) and from Mg(A):B(V)=1:2 at 600 °C (open symbols), (d) from Mg:B(II)=1:2 at 800 °C (open symbols) and 1050 °C (solid symbols), (e) from Mg:B(III)=1:2 at 800 °C (open symbols) and 1050 °C (solid symbols), (f) from Mg:B(III)=1:2 with 10% Ti at 800 °C (open symbols) and 1050 °C (solid symbols).

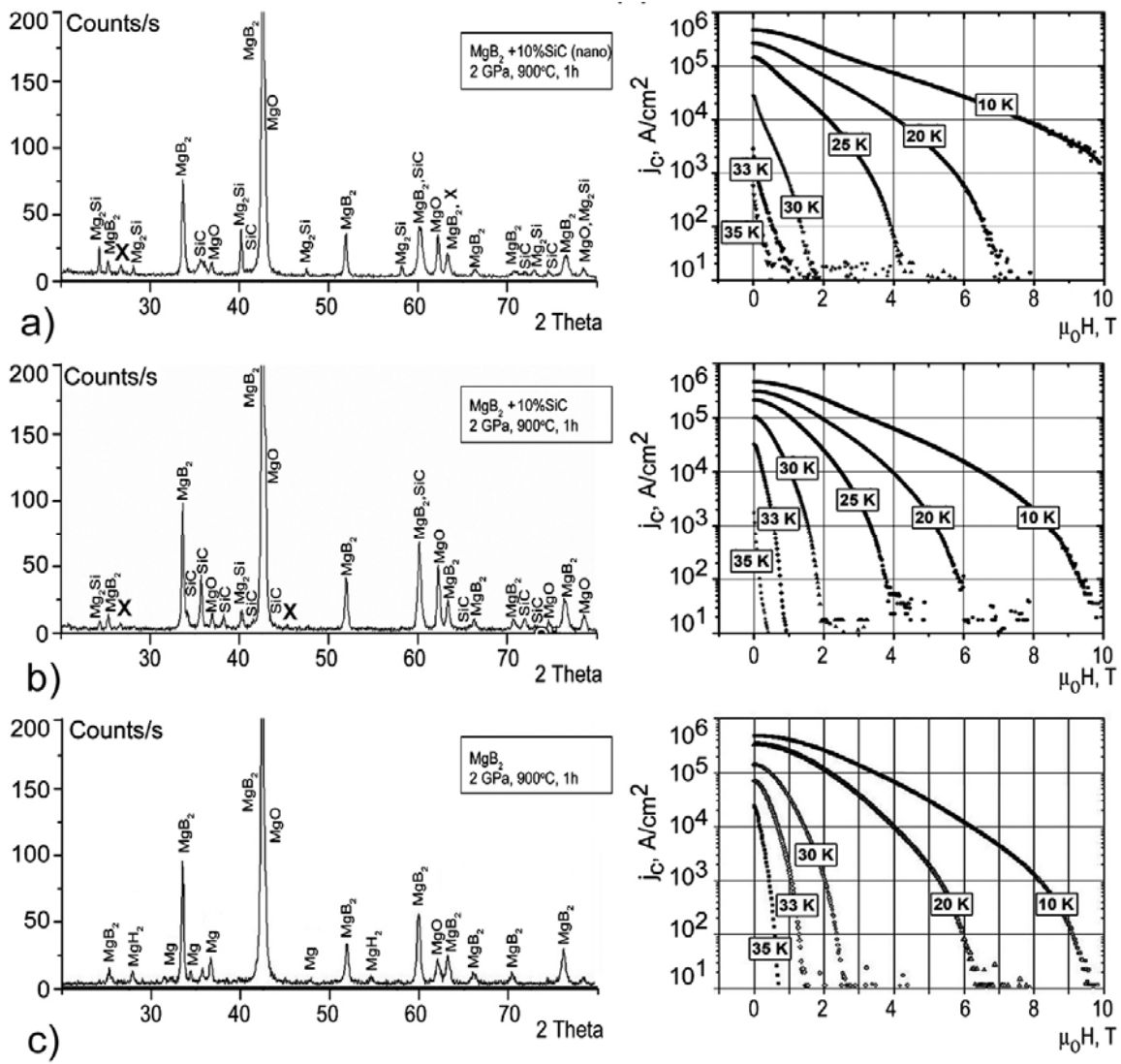

**Figure 9.** X-ray patterns and dependences of critical current density, jc, at different temperatures on magnetic field, $\mu_0 H$, for HPS-materials from Mg:B(IV)=1:2 at 2 GPa, 900 oC, 1 h: (a) with 10% nano SiC additive with particle sizes of 20-30 nm, (b) with 10% SiC additive with particle sizes of 200-400 nm, (c) without an additive.

near-$MgB_{12}$ stoichiometry, the amount of which decreases with increasing the synthesis temperature, and increases with addition of Ta or Ti. In our opinion both $MgB_{12}$ and Mg-B-O inclusions can positively affect pinning in $MgB_2$. As it was mentioned above our previous studies showed that after HP-synthesis Ta and Ti in many cases (especially when $j_c$ increased) transformed into $Ta_2H$ or $TiH_{1.924}$ hydrides and never into oxides. But up to now we have not fully understood the mechanism of Ti or Ta influence the on $j_c$ increase (compare Figs. 1a, b, 8i, f) in $MgB_2$-based materials. We can assume that the existence of the two so-called optimal temperatures for the same starting boron may be the result of competition between higher borides ($MgB_{12}$) formation and oxygen segregation to form Mg-B-O inclusions. When studying oxygen content we found that gray matrix of material high pressure synthesized at 800 °C with 10 % of Ti contained 8 % of oxygen, while gray matrix (between bright oxygen-enriched inclusions) of the materials high pressure synthesized from the same B and Mg at 1050 $^o$C with addition of 10% Ti or Ta contained only 1.5 - 5 or 6 % oxygen, respectively. So, Ti and Ta seem to promote a decrease of oxygen amount in the matrix material, thus contributing to the formation of Mg-B-O inclusions with a high oxygen content and pinning increase. The segregation of oxygen with temperature increase in materials without additions is not so pronounced.

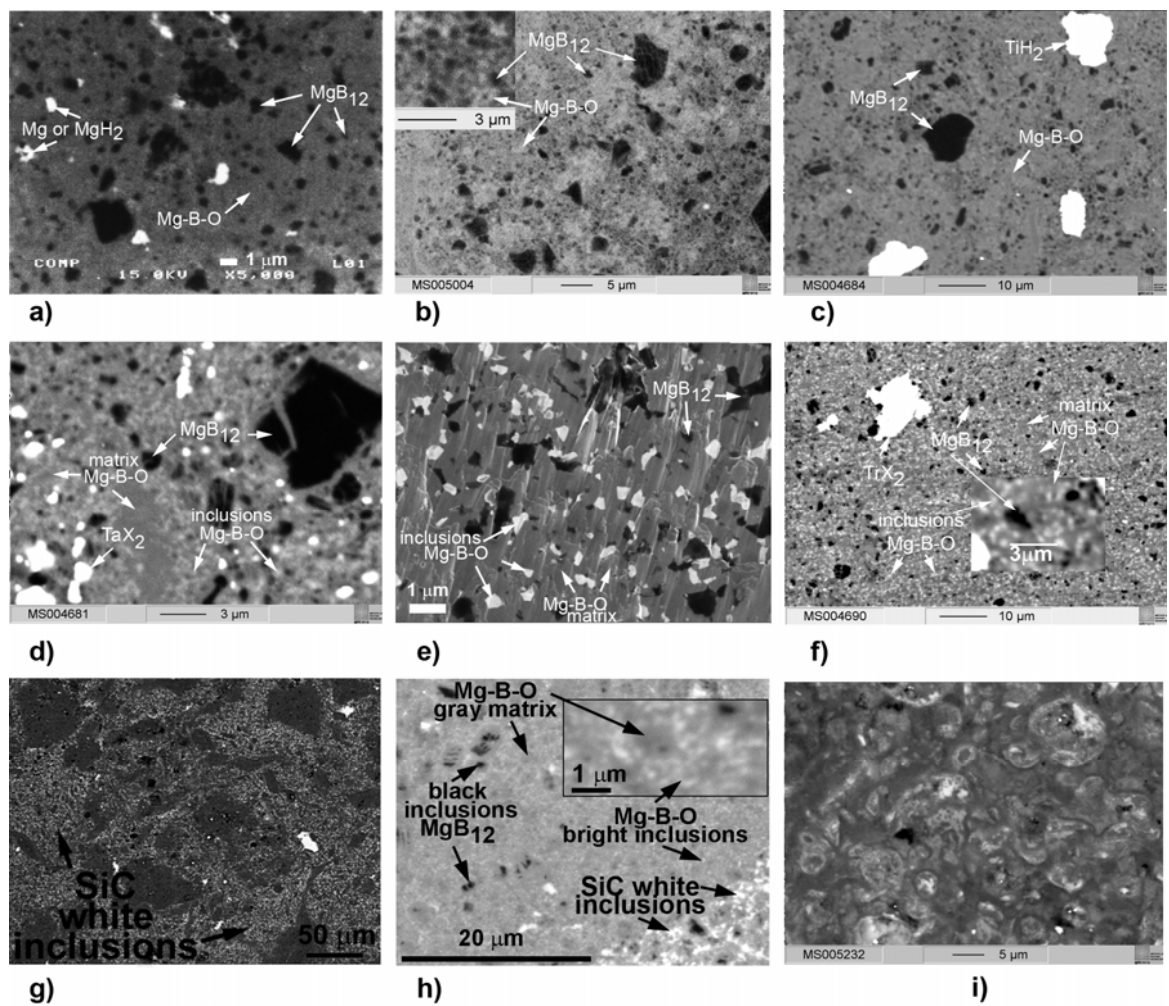

**Figure 10.** Composition images (backscattering electron images) of HPS $MgB_2$-based materials (a-h) synthesized from Mg and B taken in $MgB_2$ stoichiometry at 2 GPa for 1 h: (a) at 800 $^o$C from boron B(IV); (b) at 1050 $^o$C from boron B(III) (c) at 800 $^o$C with 10% Ti from boron B(III), (d) at 1050 $^o$C with 10% Ta from boron B(III), (e, f) at 1050 $^o$C with 10% Ti from boron B(III) under different magnifications (Fig. 10e shows the enlarged area containing no Ti , the "steers" seen in matrix are because of etching), (g, h) at 1050 $^o$C with 10% SiC (200-800 nm) from boron B(II) under different magnifications (Fig. 10h shows the enlarged area containing small amount of SiC); (i) synthesized from Mg(A) :B(V)=1:2 at 2 GPa, 600 $^o$C for 1 h.

Such high critical current densities in high magnetic fields (Figs. 8c, open symbols) exhibited by material synthesized at very low temperature 600 $^o$C (from Mg(A) powder and B(V)) were very unusual for our previous studies. The structure of this material was not very dense and rather inhomogeneous (Fig. 10 i). Up to now this material has not been studied properly, we can only assume that there is near 7 wt% of oxygen in its matrix and the composition of the matrix corresponds approximately to $MgB_3O_{0.3}$. As our former results have shown the substitution of Mg chips for 325 meshes powdered Mg did not allow us to improve $j_c$ and the use of synthesis temperature below 800 °C resulted in the absence of SC properties or in very low values of $j_c$.

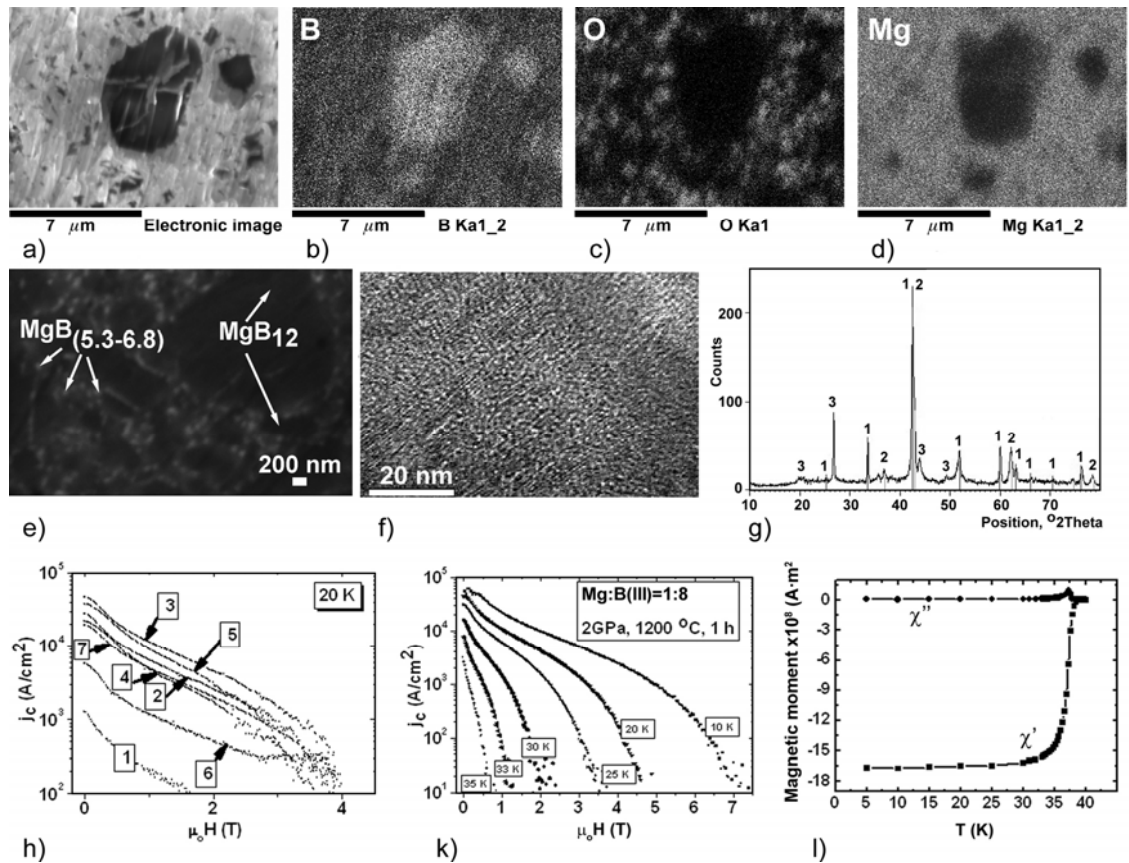

**Figure 11.** (a-d) backscattering electron image and analysis of elements distribution (the brighter looks area the higher is amount of the element under study) over the area of HPS $MgB_2$-based sample from Mg and B(III) with 10% Ti taken in $MgB_2$ stoichiometry at 2 GPa, 1050 °C, for 1 h: (a) electron image, (b - d) distribution of boron, oxygen and magnesium, respectively (the same sample is shown in Figs. 10e, f); (e, f, g, k, l) views of structure and characteristics of the material synthesized from Mg and B(III) taken into 1:8 ratio at 2 GPa, 1200 °C for 1 h: (e) backscattering SEM electron image: the stoichiometry of matrix is near $MgB_{12}$, (f) TEM image which shows the mosaic structure of a $MgB_{12}$ grain; (g) X-ray pattern of the sample (reflexes "1" and "2" coincide with those of $MgB_2$ and MgO, respectively, reflexes "3" possibly belong to $MgB_{12}$, reflex "3" at $2\Theta=27°$ coincides with that of BN), (k) dependences of critical current density, $j_c$, on magnetic fields, $\mu_oH$, (l) dependences of real ($\chi'$) and imagine ($\chi''$) parts of resistance vs. temperature; (h) dependences of $j_c$ on the external magnetic fields, $\mu_oH$, at 20K for the materials synthesized at 2 GPa for 1 h from Mg and B(III) taken in the ratio Mg:B and synthesized at $T_S$: curves 1 - 1:12, $T_S$ = 1200 °C, curve 2 – 1:10, $T_S$ =1200 °C, curve 3 - 1:8, $T_S$ =1200 °C, curve 4 - 1:6, $T_S$ = 1200 °C, curve 5 - 1:4, $T_S$ =1200 °C, curve 6 - 1:12 , $T_s$= 800 °C, curve 7-1:20, $T_s$.=1200 °C.

The notable effect of the presence of $MgB_{12}$ phase inclusions in $MgB_2$ on SC properties compeled us to focus our attention on $MgB_{12}$ synthesis. For this study we used only boron B(III) and Mg chips (the mixtures were mixed and milled in a high-speed activator). Our multitudinous studies allow us to conclude that appearance of reflex "x" (Figs. 5c, d, 9a, b) or reflexes "3" (Fig. 11g) on X-ray can be responsible for higher borides (this reflex exactly corresponds to hexagonal BN, which is used to protect samples from a graphite heater, but the samples under study contain no nitrogen). Figures

11a-e demonstrate (using backscattering electron regime) the essential difference between materials with $MgB_2$ (Figs. 11a-d) and $MgB_{12}$ (Fig. 11e) matrixes. As is shown from high-resolution TEM studies, $MgB_{12}$ grains (Fig. 11f) were far from perfection and contained many mosaic subgrains. The last mentioned reason can be an explanation to the fact why we did not get sharp reflections of Kikuchi lines from $MgB_{12}$ grains.

The specially HP-synthesized samples from mixtures with a large amount of boron (up to Mg:B=1:20) exhibited SC properties (Fig. 11h). The highest $j_c$ as well as $T_c$ near 37 K were demonstrated by samples with near $MgB_{12}$ composition of matrix (prepared from mixtures Mg:B= 1:8) whose structure and properties are shown in Figs 11e, f, g, k, l. The corrected amount of shielding fraction in the sample is 95.3% which witnesses a large volume of SC phase. As high-resolution TEM and SEM energy-dispersion analysis showed, the sample (Fig. 11e) mainly contained a phase with near $MgB_{12}$ stoichiometry and some amount of $MgB_{(5.3-6.8)}$ ($MgB_2$ was found in the form of random inclusions of less than 100 nm). A high $j_c$ were shown by the material synthesized from Mg:B=1:20 in which a high amount of SC phase also has been observed and the matrix of which as SEM study reviled had near $MgB_{12}$ composition. It should be mentioned that the materials with near $MgB_{16}$ stoichiometry of the matrix exhibited very low SC properties.

The Berkovich nanohardnes and Young modulus of the $MgB_{12}$ inclusions as estimated under the 10-60 mn-load were 32.2±1.7 and 385±14 GPa, respectively, while for sapphire 31.1±2.0 and 416±22 GPa, respectively, and for the matrix phase 17.4±1.1 and 213±18 GPa. The Vickers microharness under 4.9 N-load of HPS sample with near $MgB_{12}$ matrix (Fig. 11 f) was 25.6±2.4 GPa and with near $MgB_2$ one was 13.08±1.07 GPa. The SC properties of HPS samples with near $MgB_{16}$ matrix were very low.

So, careful structural studies using high resolution SEM and TEM, microhardness tests (which showed that $MgB_{12}$ phase is as hard as sapphire) pointed to a principal difference between samples prepared from Mg:B(III)=1:8 ratio and $MgB_2$-based materials. The foregoing allows us to conclude that in materials with near $MgB_{12}$ matrix the comparatively high critical current densities can be realized.

**4. Conclusions**

The SC properties of $MgB_2$-based materials depend to a large extend on the distribution of nanostructural inhomogenities with higher boron and oxygen contents (higher borides and Mg-B-O inclusions) as well as on the presence of hydrogen, which in turn depend to a high extent on the quality of initial boron or magnesium diboride powders and can be regulated by synthesis or sintering temperatures and by the type and amount of alloying additives such as Ti, Ta, SiC, etc.

At low synthesis temperatures (about 800 °C) the presence of Ti or Ta promotes the formation of $MgB_{12}$ inclusions, which positively affects pinning in $MgB_2$-based materials, especially in high magnetic fields. Under high synthesis temperatures (about 1050 °C) adding of Ti and Ta seems promotes a decrease of the amount of oxygen in the matrix material by contributing to the formation of inclusions with a high oxygen content (Mg-B-O – inclusions) which can act as pinning centers as well. The segregation of oxygen and formation of MgB-O inclusions with an increase of synthesis temperature in materials without additives are not so pronounced. The better SC properties can be reached when it is no notable reaction between SiC and $MgB_2$. Probably the improvement in $j_c$ in high pressure-synthesized SiC-doped material is due to the formation of nanostructural inhomogeneties in the $MgB_2$ matrix and due to the presence of dispersed small SiC grains than due to carbon incorporation in $MgB_2$ crystal structure.

There are many grounds to assume that in materials with $MgB_{12}$ matrix high SC properties can be attained as in the case of $MgB_2$ matrix.